\documentclass[10pt,letterpaper,twocolumn]{article} 

\usepackage{ol2}
\usepackage[draft]{hyperref}
\usepackage{amsmath}

\begin{document}

\twocolumn[ 

\title{Broadband light coupling to dielectric slot waveguides with tapered plasmonic nanoantennas}

\author{I. S. Maksymov,$^{1,2,^{*}}$ and Yu. S. Kivshar$^2$}

\address{
$^1$School of Physics, University of Western Australia, 35 Stirling Highway, Crawley WA 6009, Australia \\
$^2$Nonlinear Physics Centre, Australian National University, Research School of Physics and Engineering, \\ Canberra ACT 0200, Australia \\
$^*$E-mail: ivan.maksymov@uwa.edu.au, mis124@physics.anu.edu.au
}

\begin{abstract}
\noindent We propose and theoretically verify an efficient mechanism of broadband coupling between incident light and on-chip dielectric slot waveguide by employing a tapered plasmonic nanoantenna. Nanoantenna receives free space radiation and couples it to a dielectric slot waveguide with the efficiency of up to $20$\% in a broad spectral range, having a small footprint as compared with the currently used narrowband dielectric grating couplers. We argue that the frequency selective properties of such nanoantennas also allow for using them as ultrasmall on-chip multiplexer/demultiplexer devices.
\end{abstract}


 ] 

\noindent Broadband subwavelength confinement, guiding, and manipulation of light are crucial for the increase in the integration density of active and passive nanodevices for on-chip optical communication systems. One of the principal specifications for such systems is to comprise broadband coupling between the outside micro-optical devices (lasers, standard optical fibers) and integrated on-chip nano-optical devices (power splitters, optical amplifiers, optical modulators, filters, etc.). In this case, it is of vital practical importance that light should be coupled directly from free space or from an optical fiber to planar integrated on-chip components \cite{1}.

\begin{figure}[htb]
\centerline{
\includegraphics[width=8.5cm]{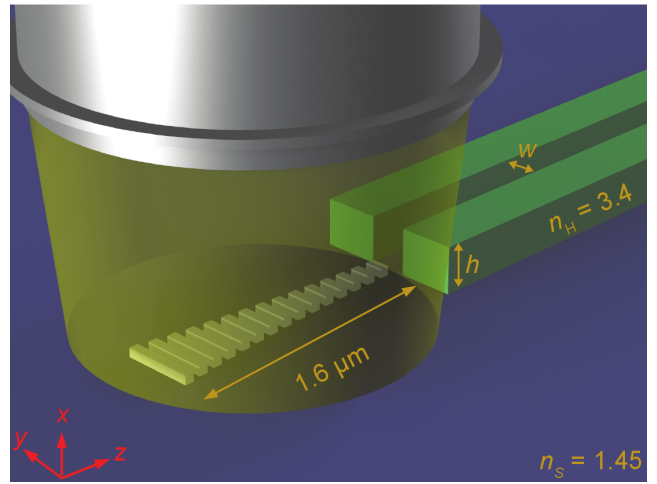}}
\caption{Broadband coupling of light from free space to an on-chip dielectric slot waveguide by means of a tapered plasmonic nanoantenna. Light {\color{black} can be focused either using a lens (shown schematically) or a lensed optical fibre}. Both nanoantenna and waveguide are placed on a dielectric substrate with low refractive index $n_{\rm{S}} = 1.45$. {\color{black} The substrate area is incomparably larger than the footprint of the nanoantenna and waveguide.} The air slot waveguide of width $w = 50$ nm is created by two high refractive index ($n_{\rm{H}} = 3.4$) bars with dimension $h = 225$ nm.}
\end{figure}

Plasmonic slot waveguides confine light in the areas of several tens of square nanometers \cite{2}. This feature makes these waveguides suitable for the use as internal on-chip optical interconnects. Light can be coupled to plasmonic slot waveguides from free space or an optical fibre. Therefore, these waveguides can also play the role of the out-of-plane  coupling interface between light and on-chip components \cite{2}. However, the small size of the slot and high reflectivity from metal layers of the plasmonic slot waveguides make it very difficult to achieve good coupling efficiencies. Several techniques were proposed to increase the coupling efficiency using, e.g., specially designed ultrasmall plasmonic nanoantennas \cite{3,4,5}. However, the practical value of the demonstrated improvements is decreased by inevitable propagation losses in the metal layers of the plasmonic slot waveguides. Furthermore, a coupler that consists of several parts increases the footprint of the device, which is highly undesirable.

Low-loss dielectric slot waveguides are reliable substitutes for the plasmonic slot waveguides \cite{6}. Such waveguides consist of a thin low-index slot embedded between two high-index bars. To fulfil the continuity condition of the normal electric displacement, a high-field concentration exists in the low-index region. For this reason the dielectric and metal slot waveguides work for the same polarization and share a similar mode profile \cite{7}. Since wave propagation is due to total internal reflection, there is no interference effect involved and the dielectric slot waveguides exhibit broadband characteristics. Most significantly, the propagation length of light confined in the dielectric slot waveguide is larger as compared to the plasmonic slot waveguide with the same dimensions. In this case, the higher propagation length is due to lower losses in the layers that form the slot. Owing to these advantages, dielectric slot waveguides can be used to improve the integration density of on-chip optical circuits.

Efficient excitation of the dielectric slot waveguide directly from free space is as challenging as the excitation of plasmonics waveguides because of a small width of the slot. It has been demonstrated so far that a dielectric slot waveguide can be excited in the vertical configuration using dielectric gratings (see, e.g., Ref.~\onlinecite{8} and references therein). The maximum theoretical coupling efficiency of about $40-60\%$ is attainable with the demonstrated devices that were designed using computer optimization. However, the total length of the dielectric gratings usually exceeds $20\mu$m, which makes them too large as compared to nanoscale integrated optical components.

In this Letter we propose and theoretically verify a broadband coupling between light and a feasible dielectric slot waveguide. The coupling is achieved using a tapered plasmonic Yagi-Uda-type nanoantenna that receives the free space radiation and couples it directly to the dielectric slot waveguide. The nanoantenna and slot waveguide {\color{black} have the dominant modes that share the same major electric field component. Since both devices also} exhibit broadband characteristics, coupling with efficiency of up to $20\%$ is demonstrated in a broad spectral range without the use of computer optimization or additional reflectors that may complicate the fabrication. The total length of the nanoantenna is below two microns, and its footprint is very small as compared to dielectric grating couplers.

\begin{figure}[htb]
\centerline{
\includegraphics[width=5cm]{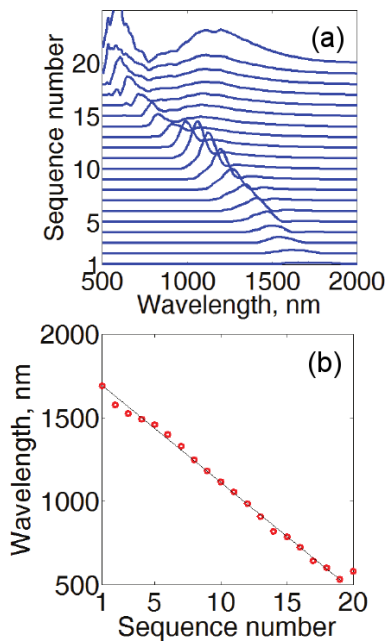}}
\caption{Operating wavelengths of the tapered nanoantenna receiving light from the dielectric slot waveguide. (a) Light spectra received by the individual elements of the nanoantenna. (b) Operating frequencies extracted from the maxima (a). Straight black line confirms the direct correlation between the operating wavelengths and the lengths of the nanoantenna elements.}
\end{figure}

A schematic of the proposed device is shown in Fig. $1$. We suggest employing a tapered plasmonic nanoantenna \cite{9} that is brought into physical contact with the air-slot dielectric waveguide. Both nanoantenna and slot waveguide are placed on the dielectric substrate with low-refractive index $n_{\rm{S}} = 1.45$. The presence of the substrate is known to decrease the forward-emitting capability of the nanoantenna because the nanoantenna emits light to the optically denser substrate \cite{9}. However, as demonstrated below, this situation changes when light emitted by the nanoantenna is coupled to the dielectric slot waveguide.

\begin{figure*}[htb]
\centerline{
\includegraphics[width=18cm]{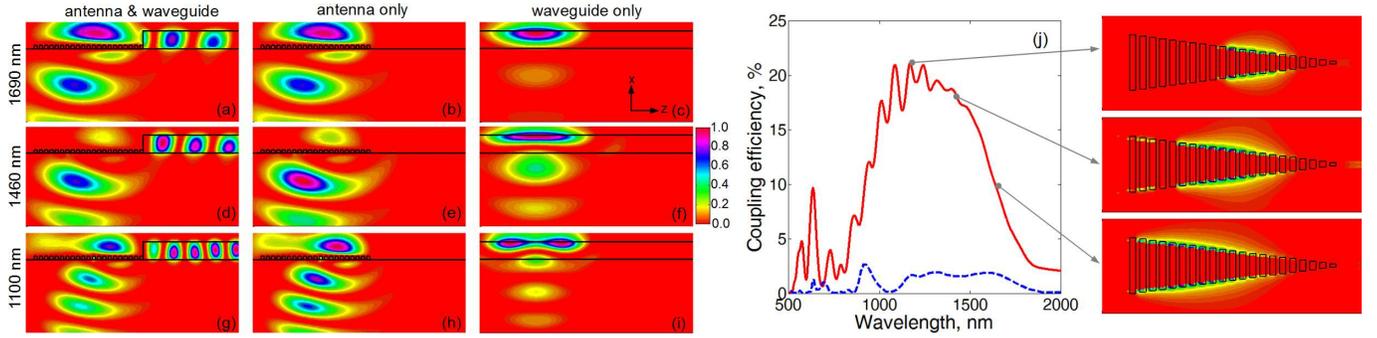}}
\caption{(a-i) {\color{black} Simulation of the light focusing from the far field on the tapered nanoantenna coupled to the waveguide (leftmost column), isolated antenna without waveguide (central column) and the waveguide without the nanoantenna (rightmost column)}. All panels show {\color{black} the intensity of the major electric field component of both nanoantenna and waveguide ($|E_{\rm{y}}|^2$)} in the \textit{xz}-plane of Fig. 1 going through the middle of the nanoantenna and the waveguide slot. {\color{black} All colormap plots are normalized to unity to allow a better comparison of the intensity profiles.} Straight lines show the contours of the nanoantenna, waveguide and substrate. For example, in panel (a) small squares denote the cross-section of the nanoantenna rods and the rectangle denotes a section of the waveguide. The space below is the substrate. (j) Extracted coupling efficiency. Dashed line: light is focused on the surface of the isolated waveguide. Solid line: light is focused on the nanoantenna coupled to the waveguide. {\color{black}$|E_{\rm{y}}|^2$} color maps on the nanoantenna surface are shown in the insets.}
\end{figure*}

In our calculations, we use the nanoantenna consisting of $21$ silver nanorods with a square $50\times50$ nm$^2$ cross-section. The spacing between the nanorods is $30$ nm. The length of the nanorods is tapered in the direction pointing to the edge of the slot waveguide. The length of the longest nanorod is $445$ nm; this nanorod operates as the reflector \cite{9}. The length of the shortest nanorod is $40$ nm, i.e. it is slightly shorter than the width of the air slot of the waveguide. The gradual length tapering from the longest to the shortest nanoantenna elements ensures broadband light focusing to the air slot of the waveguide. We note that the chosen dimensions of the nanoantenna and waveguide are attainable when using the current fabrication technologies \cite{6, 10}.

Broadband spectral characteristics of the nanoantenna coupled to the slot waveguide are investigated numerically. Due to the existence of multiple dielectric-metal discontinuities in the proposed structure, in-house finite-difference time-domain (FDTD) software \cite{11, 12} is employed. The frequency-dependent dielectric function of silver is taken from Ref.~\onlinecite{13}. The open regions of the computation domain (air layer above the nanoantenna and dielectric substrate as well as open edge of the waveguide) are truncated using perfectly matched layers (PMLs) (see, e.g., Ref.~\onlinecite{14} where good agreement with experiment is obtained using PMLs). By using PMLs, outgoing waves are artificially absorbed and only a small part of their energy is reflected back due to PMLs imperfections. By increasing the computational domain, it is additionally verified that the use of PMLs allows simulating with good accuracy virtually infinitely large dielectric substrates and long waveguides.

We excite the open end of the slot waveguide with a broadband light source with an electric field profile matching the waveguide mode, and then detect the power accepted by the individual nanorods of the nanoantenna. Detectors are placed $5$ nm off the edges of each nanorod. Figure $2$(a) shows the detected spectra as a function of the sequence number of the nanoantenna elements (from longer rods to shorter ones). The overlap of the spectra in a broad $1100$ nm range confirms that the tapered nanoantenna receives the power from the slot waveguide in a broad spectral range. By plotting the wavelengths corresponding to the maxima of the spectra and fitting the final plot with a line [Fig. $2$(b)] we reveal a direct correlation between the operating wavelengths and tapering of the nanoantenna elements. By comparing the magnitude of the resonance peaks in Fig. $2$(a) we notice two different regimes where the magnitude reaches its maximum. Higher coupled power levels at the wavelength of around $600$ nm are due to the physical proximity of the shortest nanoantenna rods (sequence numbers $15-20$) to the waveguide end. Longer wavelengths are also coupled to the nanoantenna. However, in accord to the design, the plasmon-enhancement at these wavelengths is more pronounced at longer nanorods numbers $3-12$.

In our simulations, we consider light focused from free space on the nanoantenna surface (see Fig. $1$). {\color{black} The incident light is \textit{y}-polarized and it matches the polarization of the major electric field component of the nanoantenna and waveguide modes}. The wavelength spectrum of incident light is chosen according to the total operating band of the nanoantenna [Fig. $2$(b)]. Three scenarios are considered in which (i) nanoantenna is coupled to the dielectric slot waveguide (Fig. $1$), (ii) nanoantenna sitting on the substrate is alone, and (iii) light is focused on the waveguide and the nanoantenna is removed. We conduct simulations for different dimensions of the dielectric substrate and reach good convergence results.

Figures $3$(a-i) show {\color{black}$|E_{\rm{y}}|^2$} profiles simulated for three representative operating wavelengths in Fig. $2$. A close observation shows that a significant portion of incident light is coupled to the waveguide by means of the nanoantenna (left column in Fig. $3$). The efficiency of this coupling will be discussed below. These simulations also confirm other loss mechanisms of light delivered to the nanoantenna: reflection from the nanoantenna and light scattering into the substrate. Qualitatively similar behaviour was observed at other operating wavelengths.

Furthermore, our simulations show that the isolated nanoantenna placed on the substrate scatters light away and does not re-emit it in the forward direction (central column in Fig. $3$). Recall that isolated tapered nanoantennas suspended in air preferentially emit in the forward direction \cite{9}. Our simulations also confirm low coupling efficiency between freely propagating light and the isolated dielectric slot waveguide (rightmost column in Fig. $3$). In this case, the main reasons for low coupling are (i) reflection from the high-index layers of the waveguide and (ii) small width of the air slot that constitutes from $1/30$th to $1/10$th of the wavelength of incident light. Moreover, light coupled to the high-index layers of the waveguide escapes to the low-index substrate.

The red solid curve in Fig. $3$(j) shows the coupling efficiency of freely propagating light to the dielectric slot waveguide via the tapered nanoantenna. The blue dashed curve shows the coupling efficiency to the isolated dielectric slot waveguide. The curves in Fig. $3$(j) were obtained by taking the ratio of the {\color{black} optical intensity distributed in the waveguide to the intensity supplied to the nanoantenna from free space}. First, we observe that the excitation of the dielectric slot waveguide from free space is low efficient with the coupling efficiency that does not exceed $2.7\%$. The minima in the blue dashed curve are due to Fabry-Perot oscillations in the air-waveguide-substrate interface perpendicular to the \textit{yz}-plane of Fig. $1$.

The excitation efficiency rises up when the nanoantenna is used to the maximum efficiency of $20\%$ at the wavelengths of $1100-1300$ nm. The efficiency of at least $15\%$ is also attainable in the very broad $990-1550$ nm range. As compared with the results obtained with the nanoantenna couplers for plasmonic slot waveguides, the demonstrated efficiency levels are superior \cite{3} or slightly inferior \cite{5}, correspondingly. However, our result was achieved without any reflector and/or computer optimization aimed at increasing the overall coupling efficiency by returning scattered light back to the nanoantenna. In particular, the integration of a metal mirror with the substrate \cite{5, 15} may be one of the avenues for further improvement. It is worth noting that propagation losses in plasmonic slot waveguides are higher than in dielectric slot waveguide. Therefore, power coupled to a dielectric waveguide is virtually more useful than same power coupled to a plasmonic waveguide. As compared to dielectric grating couplers the demonstrated nanoantenna efficiency is two-three times smaller. However, dielectric couplers have several orders of magnitude larger footprint, possess a very narrow operating band and also were designed using genetic optimization algorithms \cite{8}.

We note that the multiple peaks in the coupling efficiency curve of the nanoantenna are in good agreement with the maxima of the spectra in Fig. $2$(a), which were obtained by exciting the nanoantenna from the waveguide. In particular, the highest coupling in Fig. $3$(j) also corresponds to the wavelengths of around $1100-1200$ nm [see Fig. $2$(a)]. As shown in the insets to Fig. $3$(j), different parts of the nanoantenna are responsible for light reception at different wavelengths. For instance, at wavelength of $1690$ nm all elements of the nanoantenna are excited except the reflector. At $1490$ nm the reflector element and the first four nanorods operate as an effective reflector, which leads to better nanoantenna operation and increased coupling to the waveguide. For the same reason even better coupling was achieved at $1100$ nm. At shorter wavelengths, however, only the smallest elements of the nanoantenna are excited and the rest of the elements operate as a big but also highly absorptive reflector. Therefore, in contrast the prediction in Fig. $2$, in the $500-600$ nm range the coupling properties of the nanoantenna excited from the far-field zone are poor. All these results agree with our previous observations \cite{9,10}.

Another implication of the results presented in Figs. $2$ and $3$ is that the nanoantenna is coupled to the waveguide in a broad spectral range independently of whether the nanoantenna is excited from free space or waveguide. This property allows using the nanoantenna coupler in the reverse direction. In this case, we envision the use of the nanoantenna as an optical multiplexer/demultiplexer (mux/demux). The mux functionality can be achieved in the direct regime when the nanoantenna combines several input information signals into one output signal coupled to the waveguide. In the reverse regime the nanoantenna will take a multi-channel signal from the waveguide and separate those over multiple output signals. Such a mux/demux device will have a significantly smaller footprint as compared with the dimensions of the previously suggested devices based on slot waveguides (see, e.g., Ref.~\onlinecite{16}).

In summary, we have theoretically demonstrated that up to $20\%$ of freely propagating light can be coupled to a realistic on-chip dielectric slot waveguide in a broad spectral range. This effect has been achieved by employing a tapered plasmonic nanoantenna with a very small footprint as compared to bulky dielectric grating couplers currently used for vertical-to-on-chip light coupling. The capability of the nanoantenna for receiving, emitting and coupling light in a broad spectral range allows using it as a multiplexer/demultiplexer device. Since we employ no optimization, the efficiency of the suggested coupler can  be increased by reducing light leakage from the nanoantenna to the substrate and/or returning scattered light back to the nanoantenna by using additional reflectors. Moreover, the efficiency of light collection from the nanoantenna coupled to the dielectric slot waveguide can be increased by collecting light not only above the nanoantenna but also below it.

This work was supported by the Australian Research Council. Numerical simulations were conducted on the NCI National Facility in Canberra, Australia, which is supported by the Australian Commonwealth Government. ISM gratefully acknowledges a postdoctoral research fellowship from the University of Western Australia, and also thanks Mikhail Kostylev for encouragement. The authors thank Andrey Miroshnichenko for careful reading of the manuscript.


\pagebreak

\section*{Informational Page}

\end{document}